\documentclass{ws-p9-75x6-50}
\usepackage{axodraw}

\def\ie{{\it i.e.\ }}

\def\be{\begin{equation}}
\def\ee{\end{equation}}
\def\bea{\begin{eqnarray}}
\def\eea{\end{eqnarray}}
\def\ds{\displaystyle}

\def\lam{\lambda}
\def\Lam{\Lambda}
\def\Gam{\Gamma}
\def\gam{\gamma}

\def\q{{\bf q}}
\def\p{{\bf p}}
\def\k{{\bf k}}

\def\Pone{{\bf P}_1}
\def\Ptwo{{\bf P}_2}

\begin{document}

\title{Convergence of derivative expansions in scalar field theory}

\author{Tim R.\ Morris$^1$ and John F.\ Tighe$^2$}

\address{Department of Physics, University of Southampton, \\Highfield,
Southampton SO17 1BJ, UK\\ $^1$E-mail: trmorris@hep.phys.soton.ac.uk\\ 
$^2$E-mail: jft@hep.phys.soton.ac.uk}

\maketitle

\vspace{-5cm}
\begin{flushright}
SHEP 01-06
\end{flushright}
\vspace{5cm}

\abstracts{The convergence of the derivative expansion of the exact
renormalisation group is investigated via the computation of the $\beta$
function of massless scalar $\lam\varphi^4$ theory.  The derivative
expansion of the Polchinski flow equation converges at one loop for certain
fast falling smooth cutoffs. Convergence of the derivative expansion of the
Legendre flow equation is trivial at one loop, but also can occur at two
loops and in particular converges for an exponential cutoff.
}

\section{Introduction}

The derivative expansion of the effective action within the context of the
exact renormalisation group has been shown to provide an accurate
non-perturbative approximation method for scalar quantum field
theory.\cite{berv,mor:3d}  While this statement rests on empirical fact, it
is a  challenging task to prove the applicability of the derivative
expansion non-perturbatively and in all generality since it is not a
controlled expansion in a small parameter.  Rather, it results in a
numerical series since the approximation lies with neglecting higher powers
of $p/\Lam$ (where $\Lam$ is the effective cutoff and $p$ some typical
momentum) and the flow equations require the contributing typical momentum
to be of order $\Lam$. Thus it is a non-trivial question to ask whether
such an expansion converges and, if so, whether it converges to the correct
answer. Here we address this question within perturbation
theory.\cite{mortig}

\section{Wilson/Polchinski Flow Equation}

We define modified propagators $\Delta_{UV}=C_{UV}(q^2/\Lam^2)$ and
$\Delta_{IR}=C_{IR}(q^2/\Lam^2)$, where $C_{UV} (C_{IR})$ is an as yet
unspecified function acting as an UV (IR) cutoff with the properties
$C_{UV}(0)=1$ and $C_{UV}\rightarrow0$ (sufficiently fast) as
$q\rightarrow\infty$,  with $C_{IR} \equiv 1-C_{UV}$.  We write Polchinski's
version\cite{pol} of Wilson's flow equation\cite{wilson} for the individual
vertices as
\bea\label{wpfl} 
\lefteqn{{\partial\over\partial\Lam}S(\p_1,\cdots,\p_n;\Lam) = 
\sum_{\left\{I_1,I_2\right\}}
S(-\Pone,I_1;\Lam){d\over{d\Lam}}\Delta_{UV}(P_1)S(\Pone,I_2;\Lam)} 
\nonumber \\ 
& &  \hspace{3cm}-{1\over 2}\int\!\!{d^4q\over(2\pi)^4}{d\over{d\Lam}}\Delta_{UV}(q)
S(\q,-\q,\p_1,\cdots,\p_n;\Lam), 
\eea
where $I_1$ and $I_2$ are disjoint subsets of external momenta such
that $I_1\cap I_2=\emptyset$ and $I_1\cup
I_2=\{\p_1,\cdots,\p_n\}$. The sum over ${\left\{I_1,I_2\right\}}$
utilises the Bose symmetry so pairs are counted only once \ie
${\left\{I_1,I_2\right\}}={\left\{I_2,I_1\right\}}$. The momentum
$\Pone$ is defined to be $\Pone={\sum_{\p_i\epsilon I_1}}\p_i$.

\newpage

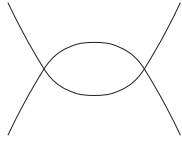
\begin{figure}[h!] 
\begin{picture}(120,50)(-10,0)
\Curve{(70,100)(80,80)(90,62)(100,46)(120,32)(134,30)
(136,30)(150,32)(170,46)(180,62)(190,80)(200,100)}
\Curve{(70,0)(80,20)(90,38)(100,54)(120,68)(134,70)
(136,70)(150,68)(170,54)(180,38)(190,20)(200,0)}
\end{picture} \hsize=5cm 
\caption{Feynman diagram contributing to four-point function 
at one loop}
\label{fig:onelp}
\end{figure}

\hspace{13pc}
\begin{minipage}[t]{15pc}
With the renormalisation condition $S(0,0,0,0;\Lam) = \lam$ imposed and the
$\beta$ function defined as $\beta(\Lam)\equiv \Lam {\partial \over \partial
\Lam}\lam$, the only contribution to the $\beta$ function at one loop takes 
the form of figure \ref{fig:onelp}.  Within the flow equation, this
contribution arises from the tree-level six-point function with two of its
legs tied together.    
\end{minipage} 

\vspace{-3.6cm}

\SetScale{0.5}
 
\noindent
The $\beta$ function at one loop is found to be 
\be\label{betawp}
\begin{array}{rl}
{\ds \beta}  & \ds =
3{\lam}^2\Lam\int\!\!{d^4q\over(2\pi)^4}  
{d \over d\Lam} \Delta_{UV}(q^2/\Lam^2)
\left[\int_{\Lam}^{\infty}\!\!\!\!\!{d\Lam_1}{d \over d\Lam_1} 
\Delta_{UV}(q^2/\Lam_1^2) \right] 
\\[3mm] & 
\ds = {6\lambda^2\over(4\pi)^2}
\sum_{n=1}^\infty{C_{UV}^{(n)}(0)\over n!}  
\int_0^\infty\!\!\!\!\!\!dx\,\,\,  x^n \,\,  C'_{UV}(x) ,  
\end{array} 
\ee \hsize=30pc
where we have taken the opportunity to perform a derivative expansion. The
derivative expansion corresponds to an expansion in the momentum dependent
part of the six-point function which 
relates to the term in square brackets in the first line of
(\ref{betawp}). This expression is evidently dependent on the exact form of
the cutoff function. Since the sharp cutoff should not be considered within
the context of the Wilson/Polchinski flow equation, we shall restrict
ourselves to smooth profiles.  With an exponential cutoff of the form
$C_{UV}(q^2/\Lam^2)=e^{-q^2/\Lam^2}$, problems arise:
\be \label{wpnocon}
\beta={6{\lam}^2\over (4\pi)^2}
\sum_{n=1}^{\infty}(-1)^{n+1}, 
\ee
Clearly this fails to converge to the correct value of $\beta={3\lam^2 \over (4\pi)^2}$, and indeed exponentials of any power (\ie
$C_{UV}(x)=e^{-x^{m}}$) suffer from the same affliction.  However there are
UV cutoffs that can provide the desired convergence.  If we consider the
much faster falling $C_{UV}(x) = \exp\left(1- e^x\right)$, we obtain the
converging series
\be
\beta={3\lambda^2\over(4\pi)^2}\left\{1.193+0-0.194-0.060+0.032 
+\cdots\right\}.
\ee

\section{Legendre flow equation at one loop}

Irrespective of the exact form of the cutoff, the only contribution to the
flow of the one-loop four-point 1PI is
\bea\label{legfl1lp}
\lefteqn{{\partial\over\partial\Lam} \Gam(\p_1,\p_2,\p_3,\p_4;\Lam)
= \int\!\!{d^4q\over(2\pi)^4}K_{\Lam}(q)}
\nonumber \\
& & \times\sum_{\left\{I_1,I_2\right\}}\Gam(\q,-\q-\Pone,I_1;\Lam)
\Delta_{IR}(|\q+\Pone|)\Gam(\q-\Ptwo,-\q,I_2;\Lam),
\eea
where the notation is the same as used in (\ref{wpfl}). Imposing the
renormalisation condition $\Gam(0,0,0,0;\Lam)=\lam$ and substituting the
classical vertex $\lam$ for the four-point functions on the right hand side
of (\ref{legfl1lp}), we find that
\be\label{betalg1lp}
\beta = 
3\lam^2\Lam\int\!\!{d^4q\over(2\pi)^4}\left({d \over d\Lam}
\Delta_{UV}(q^2/\Lam^2)\right)\Delta_{IR}(q^2/\Lam^2)={3\lam^2 \over
(4\pi)^2}. 
\ee
Note that the result is independent of cutoff function as expected, and
exact irrespective of the derivative expansion because the classical
four-point vertex carries no external momentum dependence.

\section{Sharp cutoff at two loops}

Various forms of the sharp cutoff version of the Legendre flow equation
expanded in vertices 
can be found in the literature.\cite{mor:approx,bonini,mortig}  By the
process of 
iteration, the usual Feynman diagrams can be constructed but with
restrictions on the allowed values of momentum for internal propagators.

To ensure that only renormalised quantities are used, we split the four
point function into its momentum free $[\lam(\Lam)]$ and momentum dependent
parts:\cite{bonini} 
\be\label{ren4pt}
\Gam(\p_1,\p_2,\p_3,\p_4;\Lam) = \lam(\Lam) +
\gam(\p_1,\p_2,\p_3,\p_4;\Lam),
\ee
with $\gam(0,0,0,0;\Lam)=0$.  Momentum expanding the one-loop result, we
find
\bea\label{shfl1lp}
\lefteqn{\gam(\p_1,\p_2,\p_3,\p_4;\Lam)}  \nonumber \\
& & = - \lam^2 \int_{\Lam}^{\infty}d\Lam_1
\int\!\!{d^4q\over(2\pi)^4}{{\delta(q-\Lam_1)}\over q^2}\sum_{i=2}^{4} 
\left\{{\theta(|\q+{\bf\cal P}_i|-\Lam_1)
\over (\q+{\bf\cal P}_i)^2}-{\theta(q-\Lam_1)\over q^2} 
\right\} \label{sh1lp} \\
& & = +{\lam^2\over4\pi^3}\sum_{i=2}^{4} 
\left\{{1\over6}{{\cal P}_i\over \Lam}
+{1\over720}\left({{\cal P}_i\over\Lam}\right)^3+{3\over44800}\left({{\cal
P}_i\over \Lam}\right)^5
+\cdots \right\}, \label{sh1lpans}
\eea
It is $\gam(\p_1,\p_2,\p_3,\p_4;\Lam)$ that is iterated through the flow
equation  to obtain the running of the coupling to second loop
order. 

\begin{center}
\SetScale{0.5}
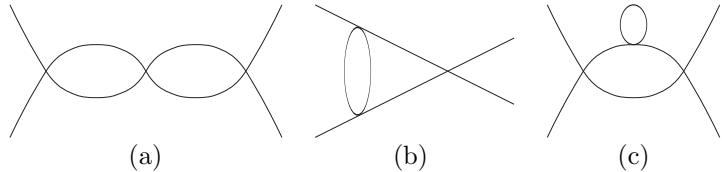
\begin{figure}[h!]
\begin{picture}(410,60)(-10,-10)
\Curve{(70,100)(80,80)(90,62)(100,46)(120,32)(134,30)
(136,30)(150,32)(170,46)(175,54)(195,68)(209,70)
(211,70)(225,68)(245,54)(255,38)(265,20)(275,0)}
\Curve{(70,0)(80,20)(90,38)(100,54)(120,68)(134,70)
(136,70)(150,68)(170,54)(175,46)(195,32)(209,30)
(211,30)(225,32)(245,46)(255,62)(265,80)(275,100)}
\put(80,-10) {(a)}
\Line(300,0)(450,75)
\Line(300,100)(450,25)
\Oval(332,50)(33,10)(0)
\put(180,-10) {(b)}
\Curve{(475,100)(485,80)(495,62)(505,46)(525,32)(539,30)
(541,30)(555,32)(575,46)(585,62)(595,80)(605,100)}
\Curve{(475,0)(485,20)(495,38)(505,54)(525,68)(539,70)
(541,70)(555,68)(575,54)(585,38)(595,20)(605,0)}
\Oval(540,85)(15,10)(0)
\put(265,-10) {(c)}
\end{picture}
\caption{Feynman diagrams contributing to the four-point function 
at two loops.}
\label{fig:twoloop}
\end{figure}
\end{center}

\vspace{-1cm}
The three possible topologies of two-loop four-point 1PI diagrams are
displayed in  figure \ref{fig:twoloop}.
Topology (a)  does not provide a
contribution to the $\beta$ function; in terms of
renormalised quantities (a) is already incorporated in the one-loop running
$\lam(\Lam)$. There are two  contributions of the form of (c), one arising
from the one-loop self energy being inserted into the one-loop four-point
function and the other from the one-loop six-point 1PI diagram with two
legs at the same vertex joined together.  It can be shown that for all
types of cutoff function, these two contributions cancel one another.
Hence the only contributions we need consider are those of topology (b).
The first comes from the iteration of the renormalised one-loop four-point
function and its contribution to the $\beta$ function is\cite{mor:approx}  
\bea\label{itsh}
&-&6 \lam \Lam \int\!\!{d^4q\over(2\pi)^4}{{\delta(q-\Lam)}\over
q^2} {\theta(q-\Lam) \over q^2}\gam(\q,-\q,0,0;\Lam) 
\nonumber \\
&=&{{\lam}^3\over (4\pi)^4} {1\over\pi}
\left(8+{1\over15}+{9\over2800}+\cdots \right).
\eea
The next two parts arise from the one-loop six-point 1PI diagram with two
legs from different vertices joined up. The first contributes
\bea\label{sh6ptsl}
&-& 6 \lam^3 \int\!\!{d^4q\over(2\pi)^4}{{\delta(q-\Lam)}\over q^2}
\int_{\Lam}^{\infty}\!\!\!\!d\Lam_1 \!\!\!
\int\!\!{d^4p\over(2\pi)^4}{{\delta(p-\Lam_1)}\over p^2}
{\theta^2(|\p+\q|-\Lam_1)\over {|\p+\q|^4}} 
\nonumber \\
&=& -12{{{\lam}^3}\over (4\pi)^4}{1\over \Lam}{1\over
\pi}
\left({\pi\over2}-{10\over9}+{\pi\over4}-{63\over
100}+{\pi\over6} +\cdots \right) ,
\eea
while the other is 
\bea\label{sh6ptqu}
&-&12\lam^3 \int\!\!{d^4q\over(2\pi)^4}{{\delta(q-\Lam)}\over
q^2}\int_{\Lam}^{\infty}\!\!d\Lam_1 
\int\!\!{d^4p\over(2\pi)^4}{{\delta(p-\Lam_1)}\over p^2}   
{\theta(|\p+\q|-\Lam_1)\,\theta(p-\Lam_1)\over {p^2|\p+\q|^2}}
\nonumber \\
&=&-12{{{\lam}^3}\over (4\pi)^4}{1\over \Lam}{1\over 
\pi}  \left({\pi\over2}-{2\over9}-{1\over300}- 
{3\over15680}+\cdots \right) .
\eea 

\begin{figure}[h]
\begin{picture}(50,50)(40,0)
\Oval(200,50)(40,40)(0)
\Line(125,50)(275,50)
\end{picture} \hsize=5.5cm
\caption{Feynman diagram contributing to wave function renormalization
at two loops.}
\label{fig:wfn}
\end{figure}
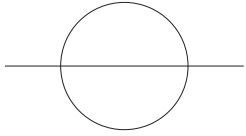

\hspace{13pc}
\begin{minipage}[t]{15pc}
While both of these series converge, the first only does so very slowly
and indeed for a $O(\partial^2)$ operator (or higher) this diagram would
diverge. 

The final contribution to the $\beta$ function at two-loop order takes
account of the wavefunction renormalisation due to figure \ref{fig:wfn}
which appears through  $\Sigma(k;\Lam)|_{O(k^2)}=[Z(\Lam)-1]k^2$.

\end{minipage}

\vspace{-4cm}

\SetScale{0.6}

\noindent
From this we find
\bea\label{shwfn}
\lefteqn{ k^2{\partial\over\partial\Lam}Z(\Lam) } \nonumber
\\
& & \left. =\lam^2\int\!\!{d^4q\over(2\pi)^4}{
{\delta(q-\Lam)}\over q^2}
\int_{\Lam}^{\infty}d\Lam_1\int\!\!{d^4p\over(2\pi)^4} 
{\delta(p-\Lam_1)\over p^2} 
{\theta(|\p+\q+\k|-\Lam_1)\over |\p+\q+\k|^2}
\right|_{O(k^2)} \nonumber
\\ 
& & = -{\lam^2k^2\over (4\pi)^4} {1\over \Lam} {1\over \pi}
\left({1\over 2}+{1\over 48}+{3\over 1280}+\cdots\right). \label{beesknees}
\eea
Altogether, these contributions   converge towards the correct two-loop
answer for the properly normalised $\beta$ function
\be\label{truebeta}
\beta(\Lam)=
\Lam{\partial\over\partial\Lam}\left({\lam(\Lam) \over
Z^2}\right)=
3{{\lam^2}\over {(4\pi)^2}} - {17\over 3}{{\lam^3}\over 
{(4\pi)^4}}.
\ee 

\section{Smooth cutoffs}

When a smooth cutoff is utilised, the Legendre flow equation takes on a
slightly different form to its sharp counterpart.\cite{mor:approx,mortig}

If we consider a power law cutoff $C_{UV}(q^2/\Lam^2)= {1 \over 1 +
(q/\Lam)^{2\kappa +2}}$ (with $\kappa$ a non-negative integer) the
derivative expansion fails at two loops.  For instance with the iteration
of the one-loop four-point function, we have the following contribution to
the two-loop $\beta$ function:
\bea\label{pow2lp}
\lefteqn{\sim {\lam^3\over \Lam^{2\kappa +3}} 
\int\!\!{d^4q}{q^{4\kappa}
\over[1+(q/\Lam)^{2\kappa+2}]^3} 
\int_{\Lam}^{\infty}\!\!{d\Lam_1\over \Lam_1^{2\kappa+3}} 
\int\!\!{d^4p} \times} 
\nonumber \\
& & \hspace{0.25in} \times {p^{2\kappa}\over[1+(p/\Lam_1)^{2\kappa+2}]^2}
\left[1- {1\over 1+(|\q+\p|/\Lam_1)^{2\kappa+2}}\right]
{1\over {|\q+\p|^2}}. 
\eea
With a derivative expansion the inner integral is expanded in terms of $q$,
but  when the power $q^{2m}$ is such that $m\ge \kappa +1$, the outer
integral fails to converge.  Hence the coefficients of the derivative
expansion are themselves infinite.

However, the situation is much better when an exponential
cutoff\footnote{Similar calculations using the 
exponential cutoff have been performed.\cite{papen}} of the form
$C_{UV}(q^2/\Lam^2) = e^{-q^2/\Lam^2}$ is used. The
renormalised one-loop four-point function is\cite{mor:momsca}
\be \label{sm1lpans}
\gam(\p_1,\p_2,\p_3,\p_4;\Lam)
 = -{{{\lam}^2}\over 2(4\pi)^2}\sum_{i=2}^{4}
\sum_{n=1}^{\infty}{(-1)^n\over (n+1)!\,n}\left({{{\cal P}_i}^2\over
2{\Lam_1}^2}\right)^{n}. 
\ee
When this is iterated through the flow equation, its contribution to the
$\beta$ function is\cite{mor:momsca}
\be\label{smit}
-12{{{\lam}^3}\over (4\pi)^4}{1\over
\Lam}\sum_{n=1}^{\infty}{{(-1)^{n}}\over 
{n(n+1)}} \left( {1\over 2^n}  - {1\over
2^{2n+1}} \right),
\ee
which can be shown to sum exactly to ${6{{\lam}^3}\over (4\pi)^4}{1\over
\Lam}[6\ln3+4\ln2-5\ln5-1]$. The analogue of (\ref{sh6ptsl}) is 
\be\label{sm6pt1}
12{{{\lam}^3}\over (4\pi)^4}{1\over
\Lam}\bigg(\ln{4\over 3} 
+ \sum_{n=2}^{\infty}{(-1)^n} 
\bigg[\,\ln{4\over 3}
-{1\over n}
\sum_{s=2}^{n}{n\choose s}{(-1)^s\over s-1}
\left\{1 - {1\over 2^{s-2}} + {1\over
3^{s-1}}\right\} \bigg]\bigg),
\ee
which numerically sums to ${{\lam^3}\over (4\pi)^4}{1\over
\Lam}(-2.45411725)$, and the equivalent of (\ref{sh6ptqu}) is 
\be\label{sm6pt2}
-24{{{\lam}^3}\over (4\pi)^4}{1\over
\Lam}\bigg(\ln{4\over 3}+ \sum_{n=1}^{\infty}{(-1)^{n}\over n(n+1)}
\left\{\left({2\over 3}\right)^n -\left({1\over 2}\right)^n  \right\}
\bigg),
\ee
which equates to $12{{\lam^3}\over (4\pi)^4}{1\over
\Lam} [9\ln3 -2\ln2 -5\ln5]$.  The flow of the wavefunction
renormalisation becomes 
\be\label{smwfn}
{\partial\over\partial\Lam}Z(\Lam) = 
{{\lam^2}\over (4\pi)^4}{1\over
\Lam}\sum_{n=2}^{\infty}{(-1)^n\over {2^n}} = 
{1\over 6}{{\lam^2}\over (4\pi)^4}{1\over\Lam}.
\ee
Gathering together these equations, we obtain fast convergence to the
correct $\beta$ function (\ref{truebeta}).

\section{Summary}

We have seen that convergence of the derivative expansion is an issue
sensitive both to the form of cutoff and to the specific flow equation that
is used. Nevertheless, the derivative expansion can converge even for
massless field theory, at least to  the two loops tested.
If the Wilson/Polchinski flow equation is employed, convergence for
the one-loop $\beta$ function can only be obtained with the use of very
fast falling smooth cutoffs. Convergence at one loop is trivial if the
Legendre flow equation is utilised. With a sharp cutoff in the Legendre flow
equation, the expansion for the two-loop $\beta$ function also converges
but clearly diverges 
for operators dependent on external momentum, whilst for a simple power law
cutoff the expansion ceases to make sense.  Finally, fast convergence of
the $\beta$ function at two loops was obtained using an exponential
cutoff.  These results thus shed light on the accuracy seen in the
non-perturbative calculations and further work along these lines could be
used to bound the accuracy and reliability of practical calculations.

\section*{Acknowledgements}

TRM and JFT thank PPARC for financial support.

\end{document}